# Modelling Rotating Detonative Combustion Fueled by Partially Pre-vaporized *n*-Heptane Sprays

Majie Zhao (赵马杰) and Huangwei Zhang (张黄伟)*

Department of Mechanical Engineering, National University of Singapore, 9 Engineering Drive 1, Singapore 117576, Republic of Singapore

## Abstract

Eulerian−Lagrangian simulations are conducted for two-dimensional Rotating Detonative Combustion (RDC) fueled by partially prevaporized *n*-heptane sprays. The influences of droplet diameter and total equivalence ratio on detonation combustion and droplet dynamics are studied. It is found that small *n*-heptane droplets (e.g. 5 μm) are completely vaporized around the detonation wave, while intermediate *n*-heptane droplets (e.g. 20 μm) are consumed in or behind the detonation wave, with the escaped ones be continuously evaporated and deflagrated. The droplet distributions in the RDE combustor are significantly affected by the droplet evaporation behaviors. Mixed premixed and non-premixed combustion modes are seen in two-phase RDC. The detonated fuel fraction is high when the droplet diameters are small or large, reaching its minimal value with diameter being 20 μm. The detonation propagation speed decreases with increased droplet diameter and is almost constant when the diameter is larger ($> 30$ μm). The velocity deficits are 2−18% compared to the respective gaseous cases. Moreover, the propagation speed increases as the total equivalence ratio increases for the same droplet diameter. It is also found that the detonation propagation speed and detonated fuel fraction are considerably affected by the pre-vaporized gas equivalence ratio. The specific impulse first decreases for cases with initial diameter less than 5 μm, then increases with droplet diameter between 5 μm and 20 μm, and finally decreases with droplet diameter larger than 20 μm.

* Corresponding author. E-mail: huangwei.zhang@nus.edu.sg. Tel: +65 6516 2557.

# 1. Introduction

Rotating Detonation Engine (RDE), as one of the pressure-gain combustion technologies, has the great potential to be commercialized. It has numerous advantages over other detonation engines (e.g. pulse detonation engines), including compact configuration, high frequency, high specific power output, and continuous existence of Rotating Detonation Waves (RDW's) [1,2]. In previous studies, gaseous fuels are mainly tested, including hydrogen and simple hydrocarbons [1–5]. However, a critical step towards practical RDE applications is to use liquid fuels, due to their higher energy density and easier storage.

Early interests in liquid fuelled RDE's date back to 1960s−70s [6], which was mainly motivated by rocket propulsion technology development for space exploration. In recent years, they have been revived, to develop pressure-gain combustion technology, and a series of laboratory-scale RDE experiments have been successfully performed with various liquid fuels. For instance, Bykovskii et al. [5,7,8] achieved two-phase Rotating Detonation Combustion (RDC) by liquid kerosene sprays. In their tests with an annular cylindrical combustor (diameter 306 mm), the oxygen-enriched air is used as the oxidant to initiate the detonation wave [5]. More recently, hydrogen or syngas is added in their experiments with a larger combustor (diameter 503 mm) using standard air and kerosene sprays [7,8]. In this combustor, Rotating Detonation Waves (RDW's) cannot be initiated without gaseous hydrogen or syngas. They systematically discussed the RDW propagation characteristics and propulsive performance of liquid kerosene fuelled RDE's [7,8], and found that stable RDW's cannot be achieved without addition of more chemically reactive reactant, like hydrogen or syngas.

In addition, Kindracki [9] investigated kerosene atomization characteristics in cold nitrogen flows by changing nitrogen velocity and fuel injection pattern of a model detonation chamber. They found that most of the droplets in their experiments have diameters of 20−40 µm, which can quickly evaporate and form a combustible mixture in the chamber. Kindracki [10] used liquid kerosene and gaseous air to study initiation and propagation of RDW's. In his work, continuous propagation of detonation wave was successfully achieved for a mixture of liquid kerosene and air with hydrogen addition, and velocity deficits of 20−25% are observed. Furthermore, based on a rocket-type combustor, Xue et al. [11] used liquid nitrogen TetrOxide (NTO) and liquid MonomethylHydrazine (MMH) as the propellants to study the RDW propagation mode under different mass flow rates and outlet structures. In the experiments of Xue et al. [11], the feasibility of RDE's with liquid hypergolic propellants was demonstrated and it was also found that two combustion patterns, i.e. single-wave mode and counter-rotating double-wave mode, were observed for the sustained RDWs.

The foregoing experimental studies have provided significant scientific insights about the two-phase RDC. However, detailed information about detonation and flow fields (e.g. droplet evaporation and fuel detonation) inside the channel are difficult to be measured. Alternatively, numerical simulation based on fully compressible reacting two-phase flows is a promising method to understand fundamental physics and assisting practical design and optimization of liquid fuelled RDE's. For instance, based on Eulerian−Lagrangian method, Sun and Ma [12] used octane and air as the reactants to numerically investigate the effects of air total temperature and fuel inlet spacing on the two-phase RDW. They found that increasing the fuel inlet spacing results in reduction of RDW propagation speed. They also observed that increasing air total temperature would increase the fuel inlet spacing limit for stable RDC. Moreover, Hayashi et al. [13] used Eulerian−Eulerian method to simulate the JP-10/air RDE's with different droplet sizes (i.e. 1−10 μm) and pre-evaporation factors (i.e. 0−100%). From their results, high liquid droplet densities are found along the contact surface between the fresh and burned gas. They also observed that the non-reactive fuel pockets behind the detonation wave and highlighted the possible detonation quenching mechanisms caused by the interactions between the detonation front and fuel droplets.

In this work, we aim to investigate the influences of liquid *n*-heptane properties (e.g. droplet diameter and pre-vaporization) on rotating detonation combustion and sprayed droplet evolutions in a modelled rotating detonation combsutor. The effects of initial droplet diameter and pre-vaporized gas equivalence ratio on detonation propagation speed, detonated fuel fraction in two-phase RDEs and its relevance to droplet properties, and the distributions of droplet diameter along the RDE chamber height will be discussed in detail. These have not been explored in the previous work (e.g. Ref. [5,7,8,12,13]) but are of great importance for utilization of liquid fuels in detonation propulsion innovations. Two-dimensional unrolled RDC model is used and the liquid fuel is injected into the chamber with a lean pre-vaprozied n-heptane / air mixture. The manuscript is organized as below. In Section 2 the computational method and the physical model are introduced. Results are presented in Section 3 and conclusions are made in Section 4.

## 2. Mathematical and physical models

### 2.1 *Governing equation*

In the present investigations, Eulerian–Lagrangian method is adopted to study the two-phase rotating detonation combustion. For the gas phase, the governing equations of continuity, momentum, energy and species mass fraction, together with the ideal gas equation of state, are solved [14]. They respectively read

$$\frac{\partial \rho}{\partial t} + \nabla \cdot [\rho \mathbf{u}] = S_m, \qquad (1)$$

$$\frac{\partial (\rho \mathbf{u})}{\partial t} + \nabla \cdot [\mathbf{u}(\rho \mathbf{u})] + \nabla p + \nabla \cdot \mathbf{T} = \mathbf{S}_F, \qquad (2)$$

$$\frac{\partial (\rho E)}{\partial t} + \nabla \cdot [\mathbf{u}(\rho E)] + \nabla \cdot [\mathbf{u}p] + \nabla \cdot [\mathbf{T} \cdot \mathbf{u}] + \nabla \cdot \mathbf{q} = \dot{\omega}_T + S_e, \qquad (3)$$

$$\frac{\partial (\rho Y_m)}{\partial t} + \nabla \cdot [\mathbf{u}(\rho Y_m)] + \nabla \cdot \mathbf{s_m} = \dot{\omega}_m + S_{Y_m}, (m = 1, \dots M - 1), \qquad (4)$$

$$p = \rho R T. \qquad (5)$$

Here $t$ is time, $\nabla \cdot (\cdot)$ is divergence operator. $\rho$ is the density, $\mathbf{u}$ is the velocity vector, $T$ is gas temperature, and $p$ is the pressure which is updated from the equation of state, i.e. Eq. (5). $Y_m$ is the mass fraction of $m$-th species, $M$ is the total species number. Only $(M - 1)$ equations are solved in Eq.

(4) and the mass fraction of the inert species (e.g. nitrogen) can be calculated from $\sum_{m=1}^{M} Y_m = 1$. $E$ is the total energy, defined as $E \equiv e + |\mathbf{u}|^2/2$ with $e$ being the specific internal energy. $R$ in Eq. (5) is the specific gas constant and is calculated from $R = R_u \sum_{m=1}^{M} Y_m MW_m^{-1}$. $MW_m$ is the molar weight of $m$-th species and $R_u = 8.314$ J/(mol·K) is the universal gas constant. The source terms in Eqs. (1) - (4), i.e. $S_m$, $\mathbf{S}_F$, $S_e$ and $S_{Y_m}$, account for the exchanges of mass, momentum, energy and species. Their expressions are given in Eqs. (12) - (15).

The viscous stress tensor $\mathbf{T}$ in Eq. (2) modelled by

$$\mathbf{T} = -2\mu \mathrm{dev}(\mathbf{D}). \tag{6}$$

Here $\mu$ is the dynamic viscosity and is dependent on gas temperature following the Sutherland's law [15]. Moreover, $\mathbf{D} \equiv [\nabla \mathbf{u} + (\nabla \mathbf{u})^T]/2$ is the deformation gradient tensor and its deviatoric component in Eq. (6), i.e. $\mathrm{dev}(\mathbf{D})$, is defined as $\mathrm{dev}(\mathbf{D}) \equiv \mathbf{D} - \mathrm{tr}(\mathbf{D})\mathbf{I}/3$ with $\mathbf{I}$ being the unit tensor.

In addition, $\mathbf{j}$ in Eq. (3) is the diffusive heat flux and can be represented by Fourier's law, i.e.

$$\mathbf{j} = -k\nabla T. \tag{7}$$

Thermal conductivity $k$ is calculated using the Eucken approximation [16], i.e. $k = \mu C_v(1.32 + 1.37 \cdot R/C_v)$, where $C_v$ is the heat capacity at constant volume and derived from $C_v = C_p - R$. Here $C_p = \sum_{m=1}^{M} Y_m C_{p,m}$ is the heat capacity at constant pressure, and $C_{p,m}$ is the heat capacity at constant pressure of $m$-th species, which is estimated from JANAF polynomials [17].

In Eq. (4), $\mathbf{s_m} = -D_m \nabla(\rho Y_m)$ is the species mass flux. The mass diffusivity $D_m$ can be derived from heat diffusivity $\alpha = k/\rho C_p$ through $D_m = \alpha/Le_m$. With unity Lewis number assumption (i.e. $Le_m = 1$), the mass diffusivity $D_m$ is calculated through $D_m = k/\rho C_p$. Moreover, $\dot{\omega}_m$ is the production or consumption rate of $m$-th species by all $N$ reactions, and can be calculated from the reaction rate of each reaction $\omega_{m,j}^o$, i.e.

$$\dot{\omega}_m = MW_m \sum_{j=1}^{N} \omega_{m,j}^o. \tag{8}$$

Also, the term $\dot{\omega}_T$ in Eq. (3) accounts for the heat release from chemical reactions and is estimated as $\dot{\omega}_T = -\sum_{m=1}^{M} \dot{\omega}_m \Delta h_{f,m}^o$. Here $\Delta h_{f,m}^o$ is the formation enthalpy of *m*-th species.

The liquid phase is modeled as a spray of spherical droplets tracked by Lagrangian method [18]. The inter-droplet interactions are neglected since dilute sprays (volume fraction < 0.001 [19]) are considered. Also, the shock-induced break-up is not considered in our studies, and its effects on RDC will be assessed in Section 3.6.3. The equations of mass, momentum and energy for the liquid phase respectively read [19]

$$\frac{dm_d}{dt} = -\dot{m}_d, \tag{8}$$

$$\frac{d\mathbf{u}_d}{dt} = \frac{\mathbf{F}_d}{m_d}, \tag{9}$$

$$c_{p,d}\frac{dT_d}{dt} = \frac{\dot{Q}_c + \dot{Q}_{lat}}{m_d}. \tag{10}$$

Here $m_d = \pi \rho_d d^3/6$ is the mass of a single droplet, where $\rho_d$ and $d$ are the droplet density and diameter, respectively. $\mathbf{u}_d$ is the droplet velocity vector, $c_{p,d}$ is the droplet heat capacity, and $T_d$ is the droplet temperature. Infinite thermal conductivity of the droplet is assumed since small droplets are investigated.

The evaporation rate, $\dot{m}_d$, in Eq. (8) is calculated with Abramzon and Sirignano model [20]

$$\dot{m}_d = \pi d \rho_f D_f \widetilde{Sh} \ln(1 + B_M), \tag{11}$$

where $\rho_f = p_S MW_m/RT_S$ and $D_f = 3.6059 \times 10^{-3} \cdot (1.8T_s)^{1.75} \cdot \frac{\alpha}{p_s \beta}$ are the density and mass diffusivity at the film [20], respectively. $\alpha$ and $\beta$ are the constants related to specific species [21]. $p_S = p \cdot exp\left(c_1 + \frac{c_2}{T_s} + c_3 \ln T_s + c_4 T_s^{c_5}\right)$ is the surface vapor pressure, with $T_S = (T + 2T_d)/3$ being the droplet surface temperature. The parameters $c_i$ are constants and can be found from Ref. [22].

The modified Sherwood number $\widetilde{Sh}$ in Eq. (11) is calculated as $\widetilde{Sh} = 2 + [(1 + Re_d Sc)^{1/3} \max(1, Re_d)^{0.077} - 1]/F(B_M)$, with the Schmidt number being $Sc = 1.0$. $F(\vartheta) = (1 + \vartheta)^{0.7} \ln(1 + \vartheta)/\vartheta$ is introduced to consider the variation of the film thickness due to Stefan flow effects [20]. Here $\vartheta$ represents the Spalding mass transfer number $B_M = (Y_{FS} - Y_{F\infty})/(1 - Y_{FS})$. $Y_{FS} = $

$\frac{MW_d X_s}{MW_d X_s + MW_{ed}(1-X_s)}$ and $Y_{F\infty}$ are the fuel vapor mass fractions at the droplet surface and gas phase, respectively. $MW_d$ is the molecular weight of the vapor, $MW_{ed}$ is the averaged molecular weight of the mixture excluding the fuel vapor, and $X_S = X_m \frac{p_{sat}}{p}$ is the mole fraction of the vapor at the droplet surface. Here $p_{sat}$ is the saturated pressure and is a function of droplet temperature $T_d$ based on Raoult's Law [23], i.e. $p_{sat} = p \cdot exp\left(c_1 + \frac{c_2}{T_d} + c_3 ln T_d + c_4 T_d^{c_5}\right)$.

Only the Stokes drag is included in Eq. (9), which is modelled as $\mathbf{F}_d = \frac{18\mu}{\rho_d d^2} \frac{C_d Re_d}{24} m_d(\mathbf{u} - \mathbf{u}_d)$ [24]. $C_d$ is the drag coefficient and estimated using the Schiller and Naumann model [25], and $Re_d \equiv \frac{\rho_d d |\mathbf{u}_d - \mathbf{u}|}{\mu}$ is the droplet Reynolds number.

In Eq. (10), $\dot{Q}_c = h_c A_d (T_g - T_d)$ denotes the convective heat transfer between gas and liquid phases. Here $A_d$ is surface area of a single droplet. $h_c$ is the convective heat transfer coefficient, and estimated using the correlation of Ranz and Marshall [26] through the modified Nusselt number, i.e. $\widetilde{Nu} = 2 + [(1 + Re_d Pr)^{1/3} \max(1, Re_d)^{0.077} - 1]/F(B_T)$, where $Pr$ is the gas Prandtl number. $B_T$ is the Spalding heat transfer number [20]. Furthermore, $\dot{Q}_{lat} = -\dot{m}_d h(T_d)$ in Eq. (3) accounts for the heat transfer caused by the latent heat of evaporation, where $h(T_d)$ is the vapor enthalpy at the droplet temperature $T_d$.

Two-way coupling between gas and liquid phases are considered, in terms of mass, momentum, energy and species exchanges. Therefore, the source terms for the gas phase equations read ($V_c$ is cell volume and $N_d$ is the droplet number in a CFD cell)

$$S_m = \frac{1}{V_c} \sum_1^{N_d} \dot{m}_d, \tag{12}$$

$$\mathbf{S}_F = -\frac{1}{V_c} \sum_1^{N_d} (-\dot{m}_d \mathbf{u}_d + \mathbf{F}_d), \tag{13}$$

$$S_e = -\frac{1}{V_c} \sum_1^{N_d} (\dot{Q}_c + \dot{Q}_{lat}), \tag{14}$$

$$S_{Y_m} = \begin{cases} S_m & \text{for the liquid fuel species,} \\ 0 & \text{for other species,} \end{cases} \quad (15)$$

where $-\dot{m}_d \mathbf{u}_d$ in Eq. (13) is the momentum transfer due to droplet evaporation.

## 2.2 *Physical model*

Figure 1 shows a two-dimensional (2D) rectangular domain, to mimic an annular RDE combustor. The soundness of 2D RDE modelling has been confirmed by numerous previous work (e.g. [14,27,28]). The length (*x*-direction) of the domain is 280 mm (the equivalent diameter is around 90 mm), whereas the height (*y*-direction) is 100 mm. These scales have been used in our previous modelling [14,28], and successful RDC is achieved. They are also comparable to those of laboratory-scale RDE combustors [29].

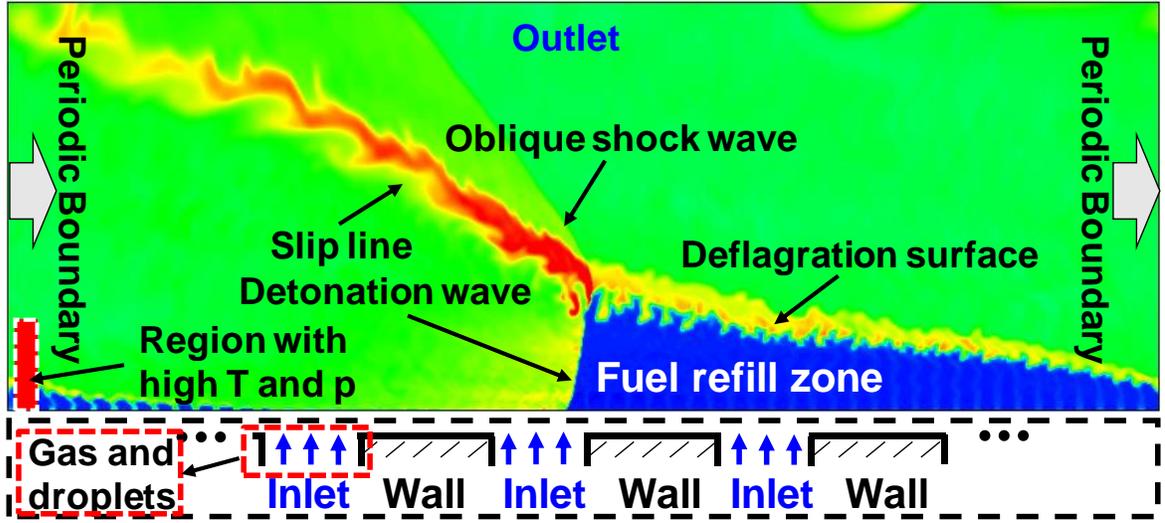

Fig. 1. Computational domain and boundary condition in two-dimensional RDE with *n*-heptane sprays. Background contour: gas temperature (200-3,000 K) of Case 2a (see details in Section 2.2). The red zone with high pressure and temperature is used for RDW initiation at *t* = 0.

The boundary conditions are marked in Fig. 1. The outlet is assumed to be non-reflective. Periodic boundaries at the left and right sides are enforced, such that the RDW can continuously propagate inside the domain. The mono-sized liquid *n*-heptane droplets are injected into the domain through 56 discrete inlets at the top head as schematically shown in Fig. 1. The initial temperature and density of the droplet are 300 K and 680 kg/m$^3$, respectively. They are carried by a premixture of air and *n*-heptane vapour, which is used to model the droplet partial pre-vaporization occurring in the upstream manifold before they enter the combustor. In the practical liquid fueled RDE experiments [3,10], it has been shown that some level of partial pre-vaporization is desirable to increase the detonability of the liquid propellant.

The wall surfaces between the discrete spray injectors (see Fig. 1) are assumed to be non-slip, impermeable and adiabatic. Here following our previous work [30], the area ratio (in 2D case, length ratio) of the injector and wall is fixed to be 2:3. Although variation in area ratio may result in different RDC behaviors (e.g. specific impulse, total pressure loss and velocity deficit) [31], these effects are not studied here.

Table 1. Liquid fuel spray information.

| Case group | Total equivalence ratio $\phi_t$ | Equivalence ratio of droplets $\phi_l$ | Initial droplet diameter $d_0$ (μm) | Initial volume fraction $\alpha$ |
|---|---|---|---|---|
| 1 | 0.8 | 0.2 | 2, 5, 10, 20, 30, 50 and 80 | 0.00014 |
| 2 | 1.0 | 0.4 | | 0.00028 |
| 3 | 1.5 | 0.9 | | 0.00064 |
| 4 | 2.0 | 1.4 | | 0.00099 |

Four case groups parameterized by total equivalence ratio, $\phi_t = \phi_g + \phi_l$, are considered, as listed in Table 1. Here $\phi_l$ and $\phi_g$ are the equivalence ratios of liquid phase and pre-vaporized premixed gas phase, respectively. Here $\phi_l$ is defined as the mass ratio of the droplets to the oxidizer normalized by the mass ratio of $n$-$C_7H_{16}$ vapor to air under stoichiometric condition. Unless otherwise stated (in Figs. 8 and 10), the equivalence ratio of the pre-vaporized premixed gas is fixed to be $\phi_g = 0.6$, whilst the liquid equivalence ratio $\phi_l$ varies from 0.2 to 1.4 in Cases 1−4, corresponding to $\phi_t = 0.8-2.0$. In each group, various initial diameters $d_0$ of mono-dispersed droplets are considered, ranging from 2 to 80 μm. Five cases from Group 2 (i.e. $\phi_t = 1.0$) with $d_0$ = 5, 10, 20, 50 and 80 μm will be studied in detail in Section 3. Hereafter, they are termed as Case 2a, 2b, 2c, 2d and 2e, respectively. The number of droplets in the whole computational domain ranges from about 40,000 to 290,000 in these cases, and the volume fractions α of the injected liquid sprays are below 0.001 (see Table 1), confirming the dilute characteristics of the sprays.

## 2.3 *Numerical implementation*

Both gas and liquid phase equations in Section 2.1 are solved by a multi-component, two-phase, and reactive solver, *RYrhoCentralFoam* [32], with two-way interphase coupling in terms of mass, momentum, energy and species (i.e. Eqs. 12-15). For gas phase, it has been validated and successfully used for gaseous supersonic flows and detonative combustion problems [30,33–35]. For liquid phase, detailed validations and verifications of the solver and sub-models are performed in our recent work [36]. Satisfactory accuracies are achieved in predicting shock wave, detonation propagation speed, and detonation cell size. More information about the numerical schemes and solution strategies can be found in Refs. [14,36,37].

For gas phase equations (Eqs. 1-4), a second-order implicit backward method is employed for temporal discretization and the time step is on the order of $10^{-9}$ s (maximum Courant number ≤ 0.1). Second-order Godunov-type upwind-central KNP scheme [38] is used for discretizing the convective terms in momentum equation, i.e. Eq. (2). To ensure the numerical stability, van Leer limiter [39] is adopted for correct flux calculations with KNP scheme. The TVD scheme is used for discretizing the convective terms in the energy and species mass fraction equations along with a second-order central differencing scheme for the diffusion terms. Computational cost associated with the latter is minimized by an efficient Operator Splitting (OS) method for both momentum and energy equations. In the first fractional step, an explicit predictor equation is solved for the convection of conserved variables (i.e. $\rho \mathbf{u}$ and $\rho E$), and in the second fractional step, an implicit corrector equation is solved for the diffusion transport for primitive variables (i.e. $\mathbf{u}$ and $E$).

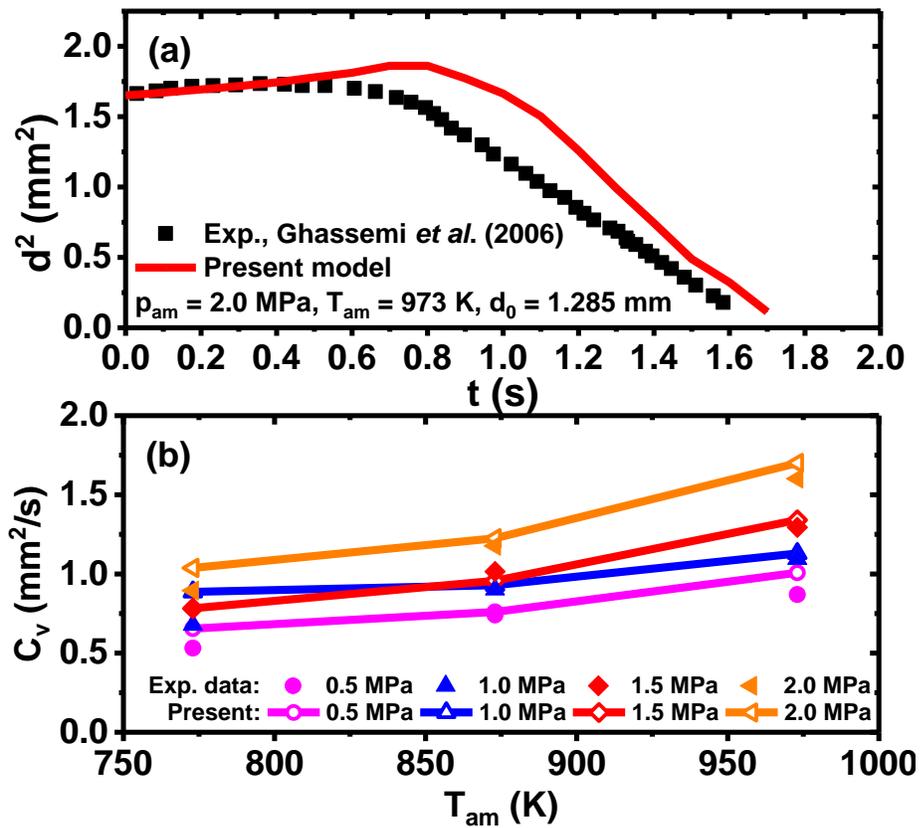

Fig. 2. (a) Time evolution of droplet diameter and (b) evaporation coefficient versus initial gas temperature. Experimental data from Ref. [40].

For the liquid phase, i.e. Eqs. (8)−(10), they are solved using first-order implicit Euler method, which is sufficiently accurate with the time step used in this work (about $10^{-9}$ s). Besides, the droplet evaporation model by Abramzon and Sirignano [20] in the *RYrhoCentralFoam* solver, which estimates the mass transfer in Eq. (8), is validated here. Figure 2 compares the diameter evolution against the measured data of single droplet (initial diameter $d_0$ = 1.285 mm) evaporation under high ambient temperature (973 K) and pressure (2.0 MPa) [40]. From Fig. 2(a), one can see that the evaporation model well reproduces the change of the droplet diameter. The duration of initial expansion, caused by the heat conduction from the surrounding gas, is slightly over-predicted. The evaporation coefficient ($C_V$, slope of $d^2 \sim t$ curve) corresponding to the steady evaporation (> 1.2 s) is reasonably predicted, i.e. 1.68, which is close to the measured value 1.60. The evaporation coefficients over a range of RDC relevant operating conditions (i.e. elevated pressures and temperatures) are further compared in Fig. 2(b) and the results demonstrate that the errors for the predicted $C_V$ are generally less than 15%. Note that extra heat conduction from the droplet holder and radiation existing in the experiments are not considered in the simulations, which may result in the foregoing deviations. Detailed discussion on the sensitivity of the simulated diameter decaying rate of single droplet to various factors are made in Ref. [33]. Validations of the same model [20] under spray detonation conditions are also available in Ref. [41]. In general, one can confirm that the accuracy of this evaporation model is reasonable in predicting droplet evaporation under detonated flow conditions.

The domain in Fig. 1 is discretized with 352,800 Cartesian cells. The cell size in the *x*-direction (RDW propagation direction) is uniform at 0.2 mm, whereas it increases from 0.1 mm at the top head (where the RDW resides) to 1 mm at the outlet in the *y*-direction. The above resolution is larger than the considered droplet diameters, which can ensure that the gas phase quantities near the droplet surfaces (critical for estimating the two-phase coupling, e.g. evaporation) can be well approximated using the interpolated ones at the location of the sub-grid droplet [41]. The mesh sensitivity will be analyzed in Section 3.6.1.

One-step reaction of 5 species (*n*-$C_7H_{16}$, $O_2$, $CO_2$, $H_2O$ and $N_2$) [42] is used for *n*-heptane detonation.

Figure 3 shows the detonation propagation speeds in gaseous mixtures of *n*-C$_7$H$_{16}$(gas) / O$_2$ and two-phase mixtures of *n*-C$_7$H$_{16}$(droplets) / air. The accuracies of the one-step mechanism are examined through comparing the results with those from a skeletal mechanism (44 species and 112 elementary reactions) [43] and experimental data [44,45]. Apparently, the results of gaseous *n*-C$_7$H$_{16}$ / O$_2$ mixture with one-step mechanism agree well with the data from the experiments and calculations with the skeletal mechanism.

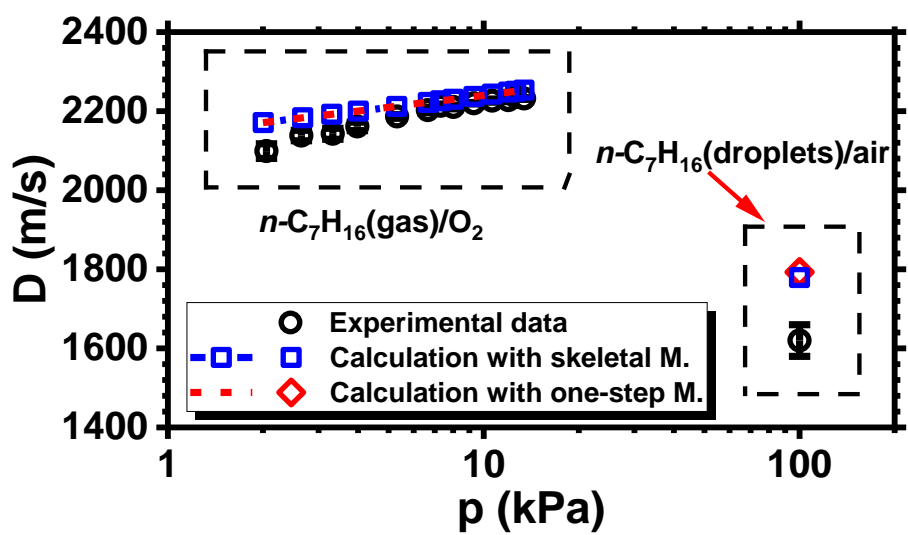

Fig. 3. Detonation propagation speeds predicted with one-step [42] and skeletal [43] mechanisms. Experimental data from Refs. [44,45].

Moreover, the detonation propagation speeds in a mist of mono-sized (5 μm) *n*-C$_7$H$_{16}$ droplets calculated with the foregoing mechanisms are very close, but both are slightly over-predicted (error 10.7%) relative to the experimental data [45]. This may be because the actual experimental conditions in gas–droplet detonations, e.g. uniformity of droplet distribution, are not quantified and therefore difficult to be fully reproduced in the simulations. In general, the accuracies of the one-step chemistry for calculating *n*-C$_7$H$_{16}$ detonation propagation are satisfactory. Further comparisons with one-step and skeletal mechanisms in 2D RDC are also made later in Section 3.6.2, including detonation propagation speed, detonated fuel fraction and mean droplet diameters.

In our simulations, a rectangular region (280 mm × 12 mm) near the inlets are initially patched with stoichiometric *n*-heptane / air mixture, whilst the rest domain is filled with air. At the left end, a rectangular hot pocket (1 mm × 12 mm, see red zone in Fig. 1) with high temperature (2,000 K) and pressure (40 atm) is used to ignite the detonation wave. In this study, the total temperature $T_o$ and total pressure $P_o$ of the injected fuels are fixed to be 300 K and 10 atm, respectively. The inlet pressure, temperature and velocity in the flow field are modelled based on the isentropic expansion relations between the top head pressure and total pressure [46,47]. Therefore, spray injection is activated if, and only if, the top head gas pressure is lower than the total pressure when the carrier gas (i.e. premixed *n*-$C_7H_{16}$/air gas) is active. Moreover, the interphase kinetic equilibrium is assumed because small droplets are considered [19], and hence the droplet injection velocity is equal to that of the carrier gas.

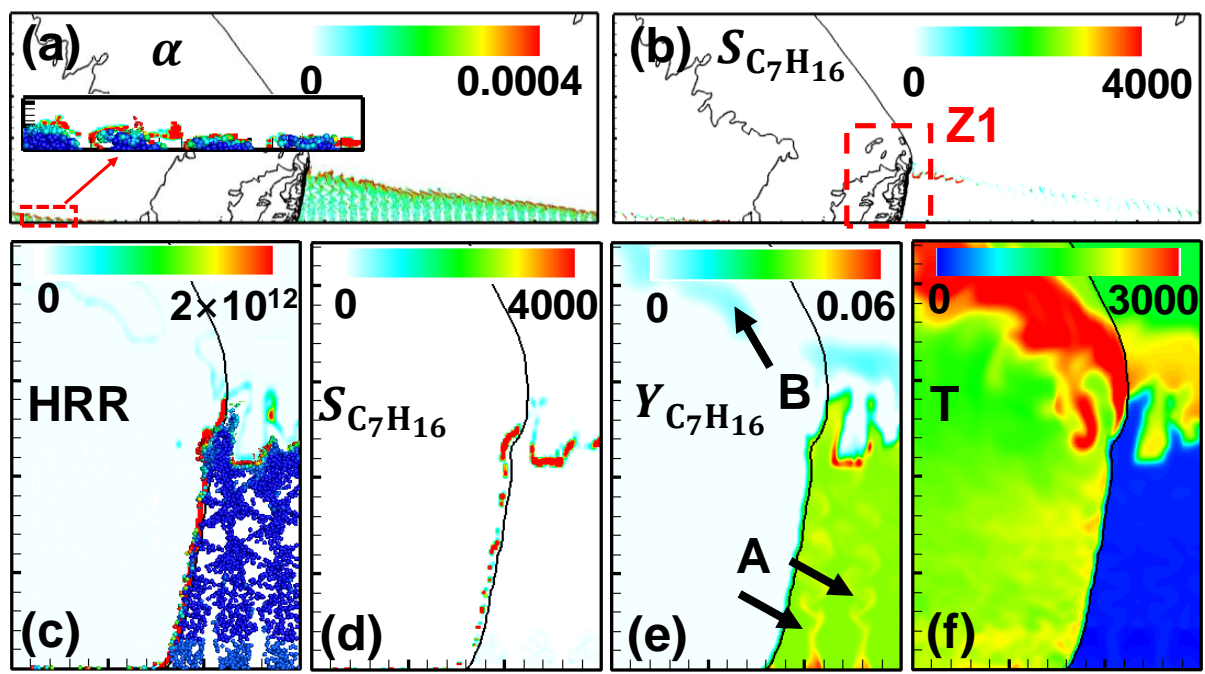

Fig. 4. (a) Droplet volume fraction $\alpha$, (b, d) evaporation rate $S_{C7H16}$ (kg/m$^3$/s), (c) heat release rate

(J/m³/s), (e) $n$-C$_7$H$_{16}$ mass fraction $Y_{C_7H_{16}}$, and (f) temperature T (K). The results from Case 2a. Black lines: pressure iso-lines of (a, b) 1–15 MPa and (c)–(f) 1 MPa. Domain size for (a, b): 280 mm × 100 mm. (c)–(f) correspond to zone Z1 in (b). Droplets colored with their temperatures are visualized in (c) and the inset of (a).

## 3. Results and discussion

### 3.1 *General feature of two-phase rotating detonations*

Figure 4 shows the distributions of droplet volume fraction α, $n$-heptane evaporation rate $S_{C7H16}$, Heat Release Rate (HRR), $n$-heptane mass fraction ($Y_{C7H16}$) and temperature ($T$) for Case 2a ($\phi_t = 1$ and $d_0 = 5$ μm). Note that $S_{C7H16}$ is the volumetric evaporation rate and predicted with Eq. (15), which will be termed as "evaporation rate" hereafter for brevity. The present simulation captures the main RDC structures, including the detonation front, oblique shock wave, and deflagration surface, as shown from Figs. 4(b), 4(c) and 4(f). They are similar to those of the purely gaseous RDC [47]. Due to the existence of dispersed droplets, new features arise. Specifically, the $n$-heptane droplets, after injected, are mainly distributed in the triangular fuel refill zone, confirmed by the high local volume fraction α in Fig. 4(a). In most of the refill zone, $n$-C$_7$H$_{16}$ mass fraction is close to that in the carrier gas, i.e. 3.82%, indicated the limited evaporation in the interior of the refill zone before the RDW approaches, because of the low local temperature (e.g. about 250 - 300 K). Nevertheless, all the droplets are vaporized along the deflagration and detonation fronts due to local high temperature, and large evaporation rate $S_{C7H16}$ can be seen there (see Figs. 4b and 4d). Therefore, almost no droplets can be found beyond the refill zone, which is also true for other simulated cases with initial droplet diameters of 2 and 5 μm. This directly leads to high $n$-C$_7$H$_{16}$ concentration around them (Fig. 4e). Ribbon-shaped zones ahead of detonation wave with locally high $n$-C$_7$H$_{16}$ concentration can be seen (see the arrows A) in Fig. 4(e), which correspond to the recirculation zone due to the walls between the spray injectors. This accumulation of $n$-C$_7$H$_{16}$ vapour results from the evaporation of droplets between the two injectors (see the inset of Fig. 4a). However, the foregoing $n$-C$_7$H$_{16}$ non-uniformity in the refill zone does not affect the RDW overall stability, and it steadily propagates at a mean speed of about 1,537 m/s. It is about 16% lower compared to the purely gaseous RDC speed, i.e. 1,830 m/s. Although the droplets are depleted inside the refill zone, however, finite $n$-C$_7$H$_{16}$ gas can still be found along the slip line (see the arrow B in Fig. 4e). Nevertheless, limited deflagrative combustion occurs there, with relatively low heat release (Fig. 4c).

Plotted in Figure 5 are the counterpart results from Case 2c with $\phi_t = 1$ but larger initial diameter, i.e. $d_0 = 20$ μm. The main RDC structures are similar to those in Fig. 4, and the RDW propagation is also stable with the speed of 1,519 m/s, about 1.2% lower than that of Case 2a. This is about 16.9% lower than the corresponding purely gaseous RDC result. However, besides in the refill zone, there are a large number of $n$-heptane droplets behind the detonation wave (see Fig. 5c), between the slip line and deflagration surface, which is characterized by the high volume fraction there. Besides droplet evaporation along the deflagrative and detonation fronts, even stronger evaporation (hence higher $S_{C7H16}$, greater than 4000 kg/m$^3$/s) proceeds immediately behind the RDW. However, local $n$-C$_7$H$_{16}$ concentration does not increase accordingly, which may be due to the distributed deflagrative combustion behind the RDW. Moreover, the mean droplet residence time in the refill zone is about 67 μs for Case 2c, much smaller than that in the whole domain (about 91 μs). However, for Case 2a with droplets fully evaporated in the refill zone, the mean droplet residence time in the refill zone is about 42 μs. Moreover, one should note that these above phenomena, e.g. the distributions of the droplets, $n$-C$_7$H$_{16}$ vapor and HRR, are also seen in the other cases with relatively larger droplet diameters.

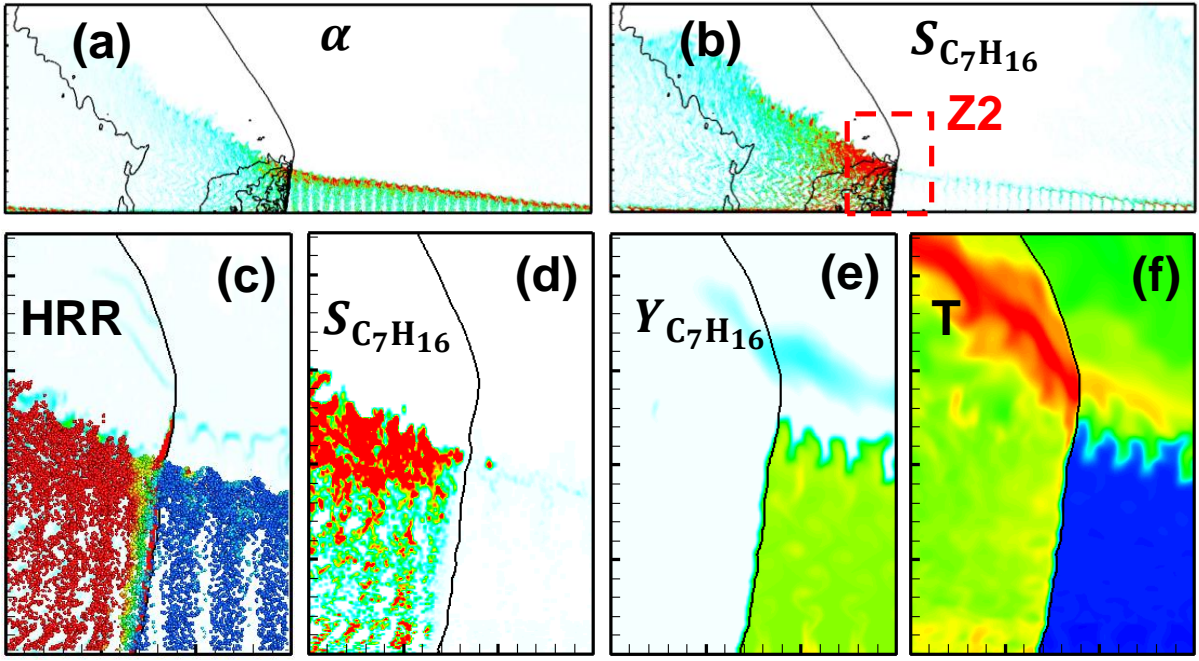

Fig. 5. (a) Droplet volume fraction $\alpha$, (b, d) evaporation rate $S_{C7H16}$ (kg/m³/s), (c) heat release rate (J/m³/s), (e) n-$C_7H_{16}$ mass fraction $Y_{C_7H_{16}}$, and (f) temperature T (K). The results from Case 2c. (c)−(f) correspond to zone Z2 in (b). Droplets colored with their temperatures are visualized in (c). Color bars same as in Fig. 4.

How the various droplet distributions affect the localized combustion mode is further elaborated with calculating the Flame Index (FI), which reads [48]

$$\text{FI} = \frac{\nabla Y_F \cdot \nabla Y_O}{|\nabla Y_F||\nabla Y_O|}, \quad (16)$$

where $Y_F$ and $Y_O$ represent the mass fractions of gaseous n-heptane and oxygen, respectively. It is often used to identify the premixed (FI = +1) and non-premixed (FI = -1) combustion modes [48]. Figure 6 shows the distributions of FI, together with HRR, for Cases 2a and 2c. One can find that the premixed combustion mode occur along the detonation and inner layer of the contact surface (see the black arrows in Figs. 6a and 6c), but non-premixed combustion is observable along the outer layer of the deflagration surface and the slip line. Recall that, for Case 2a, both n-heptane vapor and droplets are mainly consumed by the RDW (see Fig. 4). However, for Case 2c, there are still surviving droplets in the detonated gas. The *n*-heptane vapor evaporated from these droplets mixes and reacts with the oxidant due to the high pressure and temperature, leading to distributed spotty diffusion flames, as shown in Figs. 6(c) and 6(d). How the performance of the rotating detonation combustion, such as detonated fuel fraction and thrust, is affected by the diffusion combustion behind the detonation front will be discussed in Sections 3.3 - 3.4.

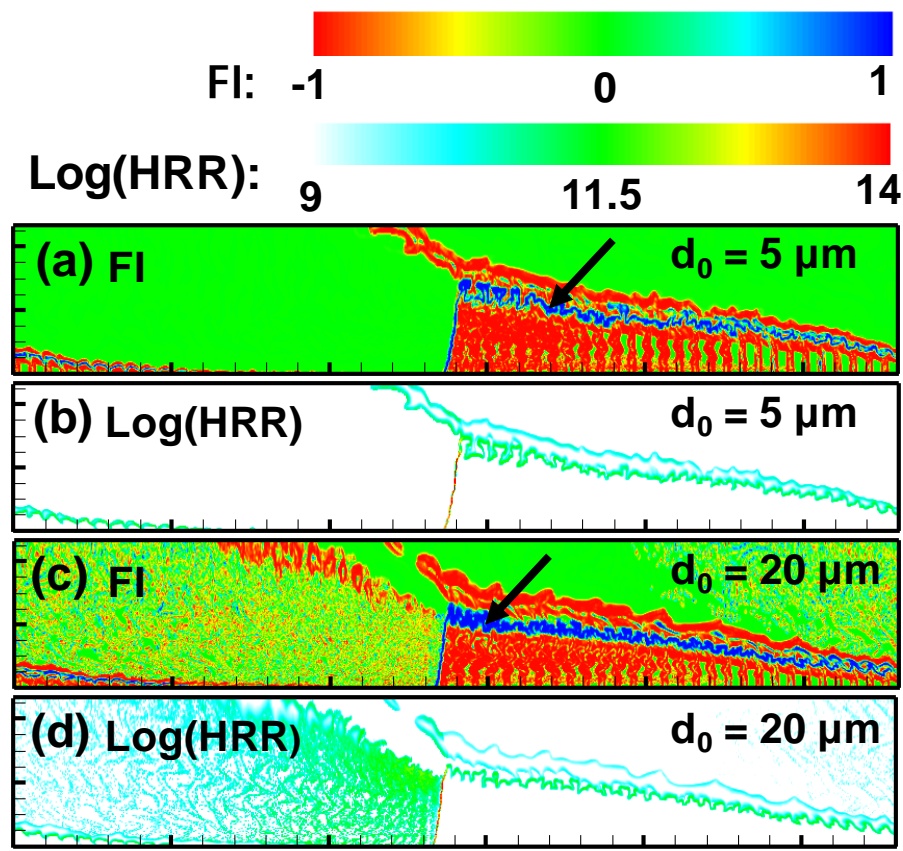

Fig. 6. (a, c) Flame index and (b, d) heat release rate (J/m³/s) for Cases 2a and 2c. Domain: 280 mm×45 mm.

## 3.2 *Droplet distribution*

Plotted in Fig. 7 are the profiles of arithmetic mean diameter ($d_{10}$) and Sauter mean diameter ($d_{32}$)

along the height of the RDE chamber. They are respectively defined as

$$d_{10} = \frac{\sum_N d}{N}, \quad (17)$$

$$d_{32} = 6\frac{\sum_N 4/3\,\pi(d/2)^3}{\sum_N 4\pi(d/2)^2}, \quad (18)$$

where $N$ is the number of droplets. Cases 2a, 2b, 2c and 2d are considered. They have the same total and pre-vaporized gas equivalence ratios (see Table 1), but different droplet diameters, i.e. 5, 10, 20 and 50 μm. In our simulations, $d_{10}$ and $d_{32}$ are estimated based on the droplets within the same height interval for all the computed instants when the RDW stabilizes. Although the mean RDW height is about 0.025 m (the dashed line in Fig. 7), however, droplets distributed with $y < 0.025$ m does not necessarily mean that they lie in the refill zone since the droplets are collected along the entire $x$ direction for a fixed height. Very close to the inlets (i.e. $y = 0$), various levels of the diameter deficit (4%−40%) can be seen for all the shown cases. This is caused by the fast evaporation of the newly injected droplets ($y \approx 0$) since they closely interact with the hot detonated gas once they are injected, as shown in the inset of Fig. 4(a).

For small droplets, e.g. $d_0 = 5$ and 10 μm, their $d_{10}$ and $d_{32}$ are almost constant for $y < 0.025$ for all the two cases, indicating limited evaporation in the refill zone. These droplets of $d_0 = 5$ μm are then fully vaporized around the detonative fronts (around 0.025 m) due to the high local temperature (see Fig. 4). However, for $d_0 = 10$ μm, residual droplets exist beyond the RDW height. They may continue evaporating along the slip line, and completely depleted around 0.05 m, around half of the RDE height.

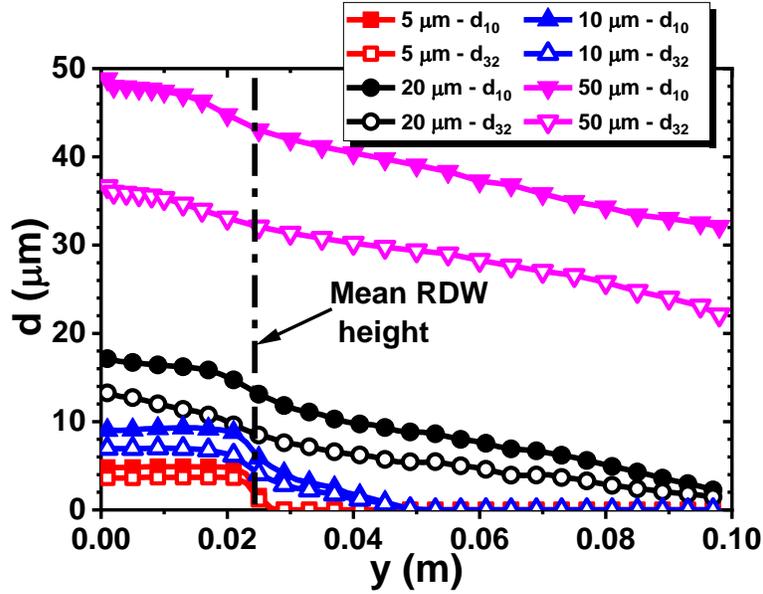

Fig. 7. Profiles of droplet diameters $d_{10}$ and $d_{32}$. Dash-dotted line: mean detonation wave height.

For $d_0$ = 20 μm, $d_{10}$ and $d_{32}$ monotonically decrease from the top head, and even at the burner exit (i.e. $y$ = 0.1 m), there are still some small-sized droplets (about 1 μm). The decrease with $y$ < 0.025 m is caused by the evaporation of the residual droplets behind the RDW. The droplets dispersed beyond $y$ > 0.025 m would continue evaporating in the detonated gas, and the vapor is burned through distributed zones, as shown in Figs. 4 and 5. When the initial droplet diameter is further increased, e.g. $d_0$ = 50 μm, although the droplets evaporate in the RDE chamber, one can see that a considerable number of large-sized droplets (around 30 μm) exists at the outlet. If the RDE is integrated with turbines, these exiting fuel droplets may affect the normal operation of the entire propulsion system [2]. Therefore, from spray RDE design perspective, besides the conventional requirements (e.g. minimum diameter) [3], critical RDE chamber heights for liquid fuel utilizations should be carefully designed to achieve better liquid fuel utilization.

3.3 *Detonation propagation speed*

Figure 8 shows the detonation wave speed as a function of initial droplet diameter $d_0$ (2−80 μm) for different total equivalence ratios $\phi_t$ (0.8−2.0). For comparisons, the results of gaseous RDC with $\phi_g = \phi_t$ (i.e. full vaporization before injection) are also plotted. In general, the wave speeds from liquid fueled RDC are 2%−18% lower than those of the corresponding gaseous RDE (open symbols in Fig. 8). These deficits are slightly lower than that reported in the liquid kerosene RDE experiments, i.e. 20−25% [49]. The velocity deficits in two-phase RDC may be associated with, e.g. non-uniform distributions (see Figs. 4e and 5e) of the dispersed droplets in the fuel refill zone, insufficient mixing between the vapor and oxidizer [10,30], and heat or momentum exchange due to the droplets near detonation front. Moreover, for the same $\phi_t$, the detonation propagation speed $D$ first decreases with initial droplet diameters. This is because smaller droplets are expected to have shorter evaporation time [40], and therefore more fuel vapor to be detonated by the RDW. For a fixed droplet diameter, larger $\phi_t$ indicates the more liquid fuel supply since the pre-vaporized $n$-$C_7H_{16}$ equivalence ratio is fixed to be 0.6, and therefore generally larger detonation propagation speed. However, when $d_0$ exceeds 30 μm, the droplet size effects on detonation propagation speed become marginal. This is reasonable because larger droplets would not markedly affect the gas composition ahead of the RDW due to their limited evaporation in the refill zone and near the RDW.

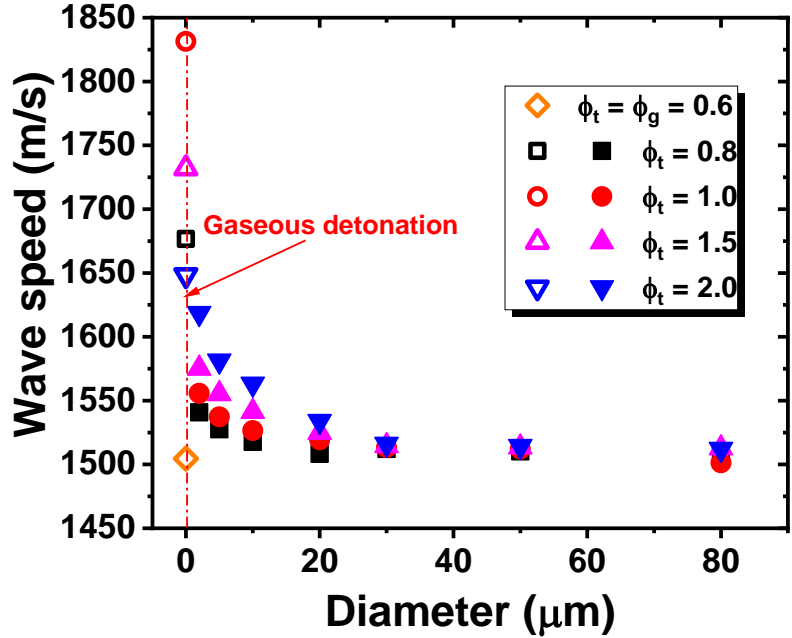

Fig. 8. Detonation wave speed as a function of initial droplet diameter and total equivalence ratio. Open symbols: gaseous RDC; solid symbols: two-phase RDC.

In Fig. 8, the equivalence ratio of pre-vaporized $n$-$C_7H_{16}$ / air mixture is fixed to be $\phi_g = 0.6$. Figure 9 further shows the detonation wave speed with variable pre-vaporized gas equivalence ratios, i.e. $\phi_g = 0.5$ and 0.8. Note that the total equivalence ratio is fixed to be $\phi_t = 1.0$ here. For a given $\phi_g$, the trends of detonation wave speed are similar to those of $\phi_g = 0.6$ shown in Fig. 8. It can also be found that the detonation wave speed increases with pre-vaporized gas equivalence ratios $\phi_g$ with the same initial droplet diameter $d_0$. The wave speeds with different $\phi_g$ and $d_0$ are always much smaller (6%−24%) than that of purely gaseous RDW with the same total equivalence ratio, i.e. $\phi_t = 1.0$ for these cases, due to the heterogeneous distributions of the gas reactants and liquid droplets in the fuel refill zone (see Figs. 4 and 5).

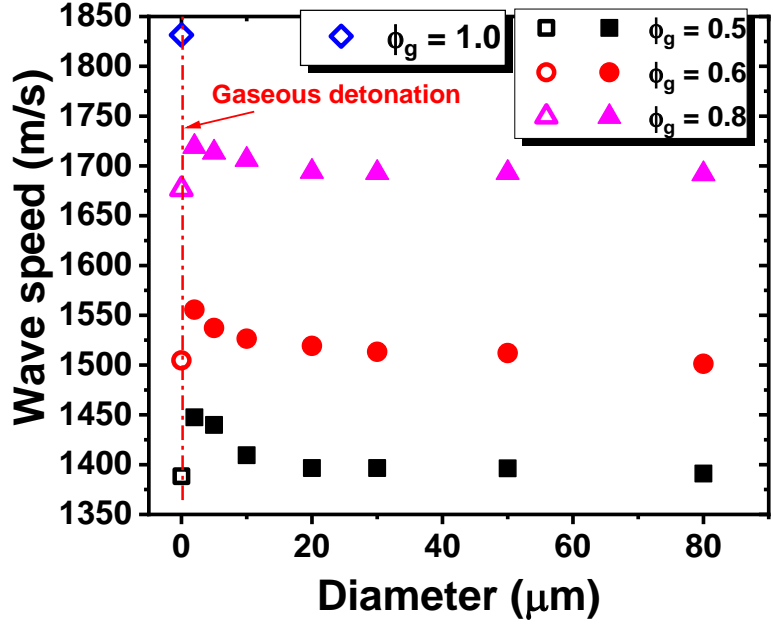

Fig. 9. Detonation wave speed as a function of initial droplet diameter and pre-vaporized gas equivalence ratio. Open symbols: gaseous RDC; solid symbols: two-phase RDC.

3.4 *Detonated fuel fraction*

Figure 10 shows the detonated fuel fraction $\psi$ under different initial droplet diameters $d_0$ and total equivalence ratios $\phi_t$. Here the detonated fuel fraction is calculated as

$$\psi = \frac{\int_V Dt_{C_7H_{16}} dV}{\int_V Dt_{C_7H_{16}} dV + \int_V Df_{C_7H_{16}} dV}, \tag{19}$$

where $Dt_{C_7H_{16}}$ and $Df_{C_7H_{16}}$ are the detonated $n$-C$_7$H$_{16}$ and deflagrated $n$-C$_7$H$_{16}$, respectively, and $V$ represents the volume of the whole computational domain. Note that the considered $n$-heptane include those from both the pre-vaporization and *in-situ* droplet evaporation in the RDE chamber. The detonated fuel fraction $\psi$ is estimated based on the volume-averaged $n$-C$_7$H$_{16}$ consumption rates conditioning on HRR greater than $10^{13}$ J/m$^3$/s, which deemed detonative combustion [14]. This critical HRR value is selected based on stand-alone numerical tests of one-dimensional detonation propagation and was also used for identifying RDW's in our previous RDE simulations [14]. Note that the errors of the results are less than 4% if this threshold value is slightly decreased or increased, e.g. $5\times10^{12}$ and $2\times10^{13}$ J/m$^3$/s.

Overall, when the droplet diameter is small, i.e. $d_0 < 20$ μm, $\psi$ decreases with increased $d_0$. Small droplets can be fully vaporized around the detonation front, thereby a higher $\psi$. When $d_0 > 20$ μm, increasing $d_0$ leads to high $\psi$. As seen in Fig. 5, large droplets continue evaporating in the denoted gas. However, with increased droplet size, the droplets take longer time to finish the evaporation and release the vapor, and therefore their kinetic effects on the RDC become weaker. The droplets with large $d_0$ cannot be fully vaporized in the RDE chamber and then exit through the outlet, and the spotty diffusion combustion behind RDW is weakened due to less fuel vapor. As such, detonated fuel fraction increases again with droplet diameter. $d_0 = 20$ μm is a critical diameter corresponding to a minimal detonated fuel fraction based on our simulations.

The detonated fuel fractions ($\psi$) with different pre-vaporized gas equivalence ratios, i.e. $\phi_g = 0.5$, 0.6 and 0.8, are also calculated and shown in Fig. 11. Here the total equivalence ratio $\phi_t$ is 1.0. For a given $\phi_g$, the trends of $\psi$ are also similar to those of $\phi_g = 0.6$ shown in Fig. 10. The critical diameter corresponding to the minimum detonated fuel fraction increases with $\phi_g$, i.e. about 10, 20 and 30 μm for $\phi_g = 0.5$, 0.6 and 0.8, respectively. Moreover, when the initial droplet diameter $d_0$ is smaller than the critical diameter, increased gas equivalence ratio would lead to high $\psi$. However, for $d_0$ greater than the critical diameter, $\psi$ of $\phi_g = 0.5$ is higher than that of $\phi_g = 0.6$, but close to that of $\phi_g = 0.8$. This may be due to the less deflagrated vapor with lower detonation temperature of $\phi_g = 0.5$ than that of $\phi_g = 0.6$.

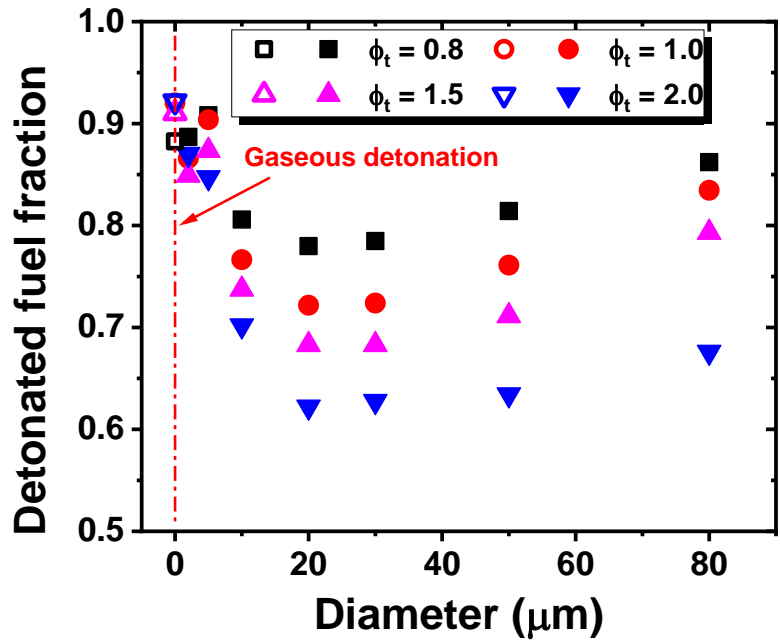

Fig. 10. Detonated fuel fraction as a function of initial droplet diameter and total equivalence ratio. Open symbols: gaseous RDC; solid symbols: two-phase RDC.

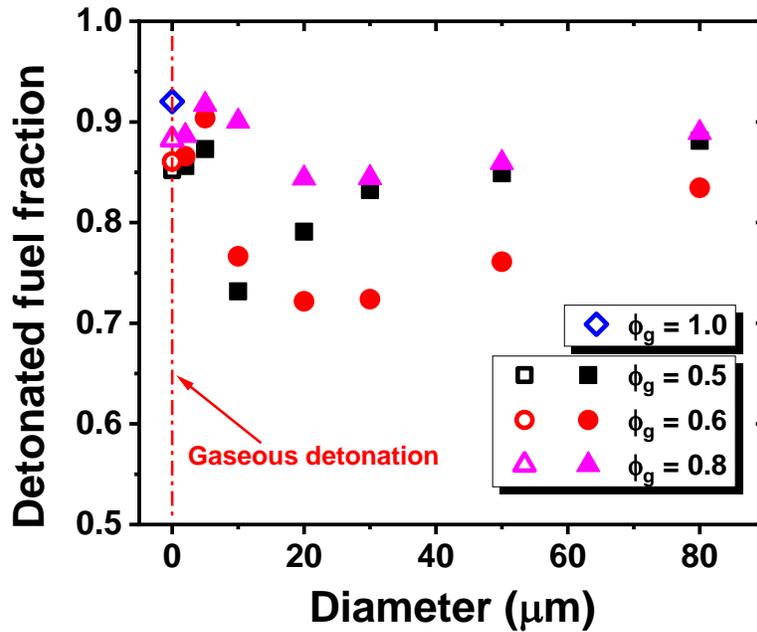

Fig. 11. Detonated fuel fraction as a function of initial droplet diameter and pre-vaporized gas equivalence ratio. Open symbols: gaseous RDC; solid symbols: two-phase RDC.

## 3.5 Propulsion performance

In order to evaluate the overall propulsion performance of rotating detonation combustion with sprayed liquid fuels, the specific impulse ($I_{sp} = \int_{A_o}[\rho u^2 + (p - p_b)]dA_o/g\dot{m}_F$) is analyzed in this section. Here $A_o$ is the area of the outlet, $u$ is the gaseous velocity at the outlet, $\dot{m}_F$ is the mass flow rate of fuel (including the gaseous and liquid fuel), $g$ is gravity acceleration, $p$ is the local pressure at the outlet and $p_b$ is the backpressure (taken as 1 atm in this work).

Figure 12 shows the specific impulse $I_{sp}$ as a function of initial droplet diameter and total equivalence ratio. Note that the pre-vaporized gas equivalence ratio is $\phi_g$ = 0.6 for all the cases in Fig. 12. For $\phi_t$ = 0.8 and 1.0, they have close specific impulses for a given droplet diameter. However, with $\phi_t$ = 1.5 and 2.0, their specific impulses are significantly reduced. This may be because the fuels from the pre-vaporization and droplet evaporation are completely reacted for $\phi_t \leq 1$, but only part of the fuel is consumed with $\phi_t > 1$ due to deficient oxidizer.

The dependece of specific impulse $I_{sp}$ on initial droplet diameter $d_0$ is *L*-shaped. Specifically, for a given total equivalence ratio, the case with purely gaseous fuel injection (open symbols in Fig. 12) has the greatest specific impulse, compared to those in two-phase scenarios, as shown in Fig. 12. This may be because a considerable part of the liquid fuel is deflagrated after evaporation along the contact surface and behind the detonation front (see Figs. 4 and 5). Also, with increased droplet diameter, the specific impulse first decreases with $d_0$ < 5 μm, then increases with $d_0$ between 5 μm and 20 μm, and finally decreases with $d_0$ > 20 μm. Although there is a peak specific impulse when the droplet diameter is around 20 μm, it has the smallest detonation fraction (see Fig. 10). This indicates that the specific impulse is affected by the deflagration and detonation as the droplet diameter changes, which will be further discussed in the following.

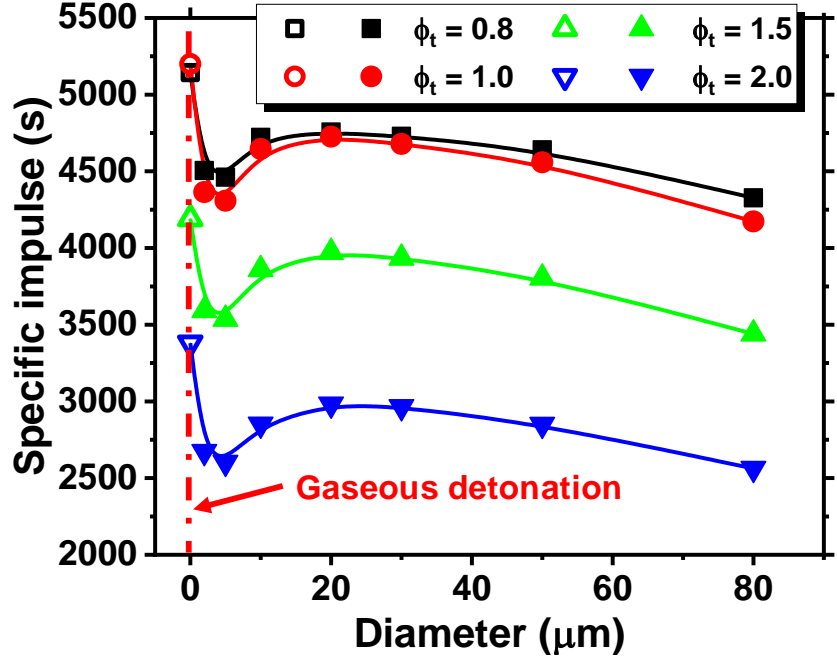

Fig. 12. Specific impulse as a function of initial droplet diameter and total equivalence ratio. Open symbols: gaseous RDC; solid symbols: two-phase RDC.

The two-phase rotating detonations with total equivalence ratio $\phi_t = 1.0$ are seclected to further articulate the droplet diameter effects on propulsion performance. Figure 13 shows the thrust force from the kinetic energy and pressure gain, as well as the powers from deflagrative and detonative combustion. The thrust force from kinetic energy is defined as $F_u = \int_{A_o} \rho u^2 dA_o$, whereas the thrust force from pressure gain is defined as $F_p = \int_{A_o} (p - p_b) dA_o$. The detonation power is estimated from $W_{Dt} = \int_V HRR_{Df} dV$, whereas the deflagration power is $W_{Df} = \int_V HRR_{Df} dV$. $HRR_{Dt}$ is the heat release rate associated with detonative combustion ($\geq 10^{13}$ J/m³/s), whereas $HRR_{Df}$ is the heat release rate in deflagrative combustion ($< 10^{13}$ J/m³/s) [28].

Since the RDE mainly utilizes the high-efficiency detonative combustion to sustain the thrust, the dominant thrust force at the outlet is from the pressure gain, which can be clearly seen in Figs. 13(a) and 13(b). When the droplet diameter is less than 5 μm, the thrust force from pressure gain $F_p$ as well as the detonative combustion decreases significantly with the droplet diameter, as shown in Fig. 13(a), leading to reduced specific impulses (see Fig. 12). One can also see from Fig. 13(b) that the detonation generated power also decreases significantly with the droplet diameter, while little change is observed for the deflagration generated power for the cases with $d_0 < 5$ μm. Also, note that the fuel droplets are completely consumed with $d_0 < 5$ μm when crossing the detonation front (see Fig. 4). This suggests that the detonative combustion would be significantly affected by small droplets ($d_0 < 5$ μm), and their effects may be mainly from the two-phase interaction close to the detonation front, such as vapor addition, energy exchange due to droplet evaporation, and momentum exchange. However, for $d_0 > 5$ μm, it is observed that the thrust force from the kinetic energy, $F_u$, first increases and then decreases with the droplet diameter, while the thrust force from pressure gain, $F_p$, changes little. This is because when $d_0 > 5$ μm considerable fuel droplets burn behind the detonation front (see Figs. 5 and 6), and therefore the deflagration power is significantly affected (see Fig. 13b).

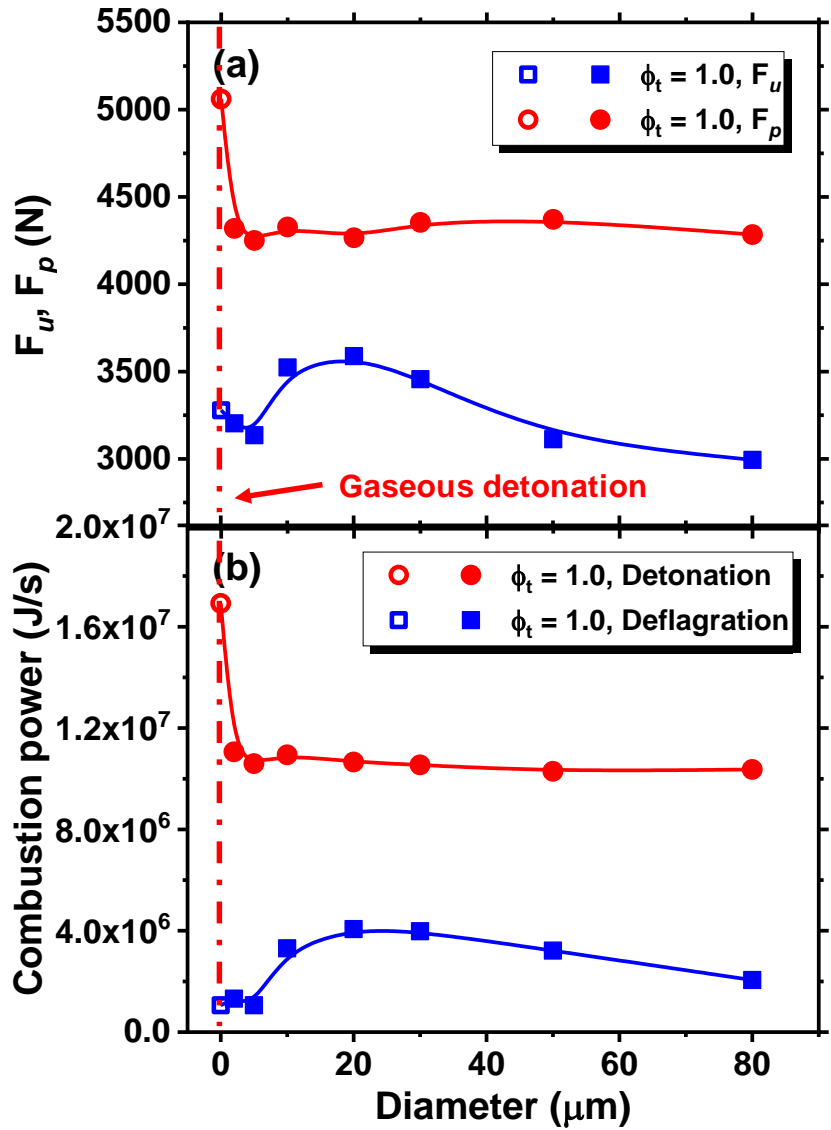

Fig. 13. (a) Thrust force from kinetic energy and pressure gain and (b) deflagrative and detonative combustion power (J/s). Total equivalence ratio is 1.0. Open symbols: gaseous RDC; solid symbols: two-phase RDC.

## 3.6 *Discussion*

To examine the effects of mesh resolution, reaction mechanism and droplet break-up on predictions of two-phase rotating detonation combustion, extra numerical experiments are performed in this section.

*3.6.1 Mesh resolution*

Simulations of Cases 2a ($d_0$ = 5 μm) and 2c ($d_0$ = 20 μm) are repeated with a finer mesh, for which the cell numbers in both *x*- and *y*-directions are doubled. For brevity, the original mesh used for the above analysis is referred to as M1, and the refined one is M2.

Figure 14 shows the time history of RDW propagation speed, and the instantaneous propagation speed is estimated based on two adjacent time instants (i.e. $\Delta t = 1 \times 10^{-5}$ s), and the detonation wave position is defined as the point with maximum heat release rate. Due to the non-uniform distributions of droplets and reactants in the fuel refill zone (see Figs. 4 and 5), all their instantaneous RDW propagation speeds slightly fluctuate. However, their mean propagation speeds and their Root-Mean-Square (RMS) values predicted with M1 and M2 are quite similar for both Cases 2a and 2c. For instance, in Case 2a, the mean RDW speeds from M1 and M2 are 1,537 m/s and 1,540 m/s, respectively, and the corresponding RMS values are 71 m/s and 69 m/s, respectively. Moreover, the instantaneous and stochastic differences in RDW propagation speed for two meshes are probably from local resolution differences in the predicted gas phase [28,50].

Moreover, to examine how the mesh resolution affects the interactions between gas and liquid phases, Fig. 15 shows the mean evaporation rate for Cases 2a and 2c predicted with M1 and M2. They are calculated based on the volume-averaged evaporation rate (i.e. Eq. 15) in the computational domain. It is seen that the mean evaporation rates are almost not affected by the mesh size. Similar observations are also made for the mean exchange rates of energy and momentum (results are not shown here). Therefore, in general, the above discussion about two-phase RDC have low sensitivity to the mesh resolution.

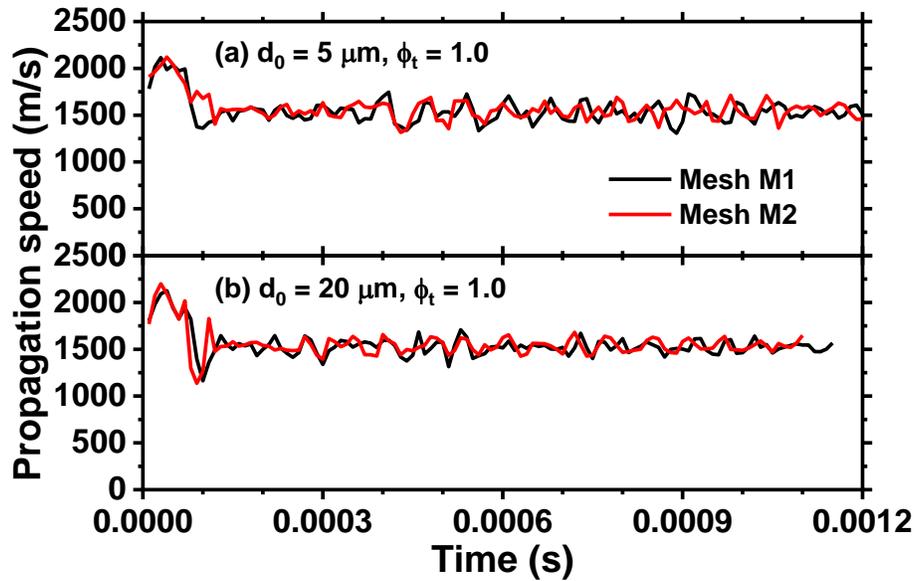

Fig. 14 Time history of detonation propagation speed in (a) Case 2a and (b) Case 2c.

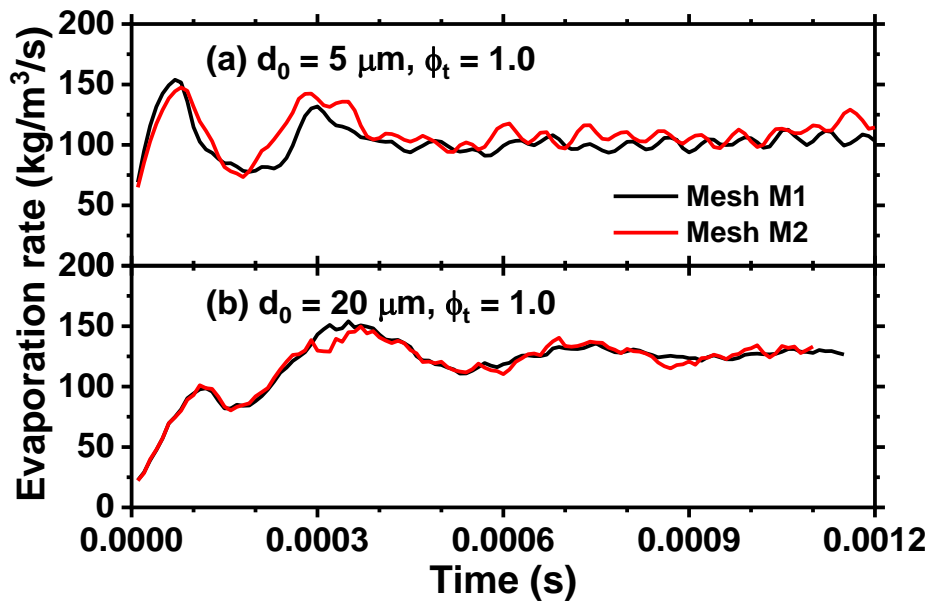

Fig. 15 Time history of mean evaporation rate in (a) Case 2a and (b) Case 2c.

*3.6.2 Chemical mechanism*

Figure 16 shows the RDW propagation speed and detonated fuel fraction as functions of initial droplet diameter with one-step [42] and skeletal mechanism [43]. The results with one-step mechanism are very close to those from the skeletal mechanism. For instance, the stoichiometric purely gaseous detonation propagation speeds with the one-step and skeletal detailed mechanisms are 1,830 m/s and 1,807 m/s, respectively, which are close to the Chapman–Jouguet speed, i.e. 1,850 m/s. Moreover, for two-phases case with different droplet diameters (i.e. 5, 20 and 80 μm), the differences of the detonation propagation speeds are less than about 2.5%. In addition, the detonated fuel fractions from the skeletal mechanism is slightly lower than those from one-step mechanism. This is probably because more deflagrations along the contact surface are captured with the skeletal mechanism [14].

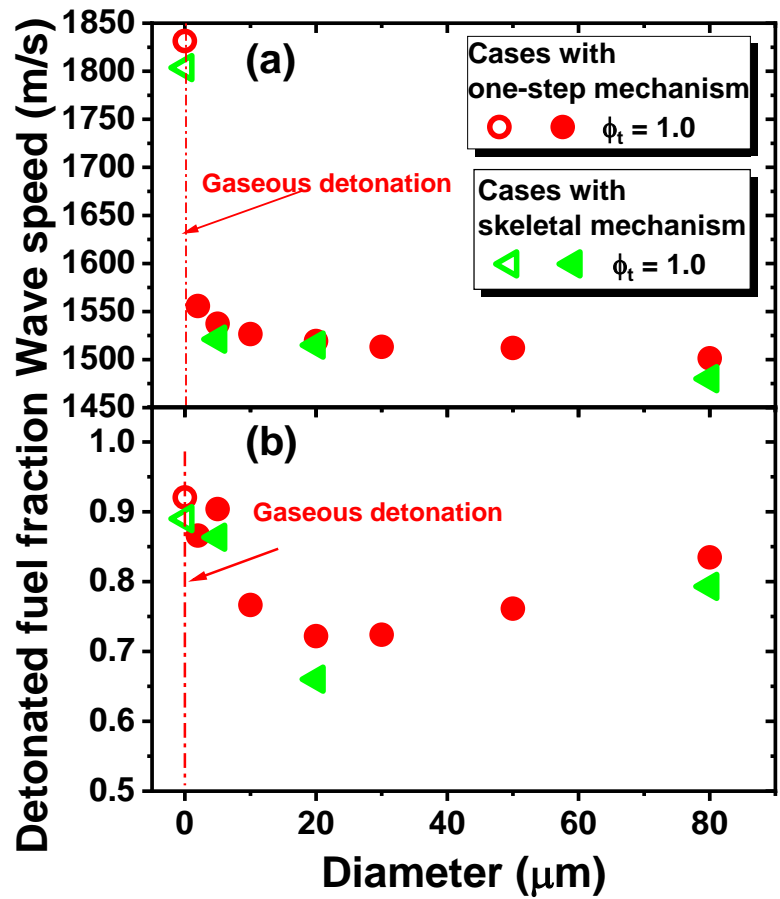

Fig. 16. (a) Detonation wave speed and (b) detonated fuel fraction as functions of initial droplet diameter. Open symbols: gaseous RDC; solid symbols: two-phase RDC.

Fig. 17 compares the profiles of the arithmetic mean diameter ($d_{10}$) and Sauter mean diameter ($d_{32}$) in Cases 2a and 2c predicted with the two mechanisms. It is shown that the droplet diameters decay more quickly when the skeletal mechanism is used. This may be caused by the continued gas phase chemical reaction behind detonation wave due to the intermediate species, leading to local heat release and hence enhanced droplet evaporation rate. In general, although there are some localized sensitivities to reaction mechanism, the accuracy of one-step chemistry in modelling liquid fuel RDC can be confirmed.

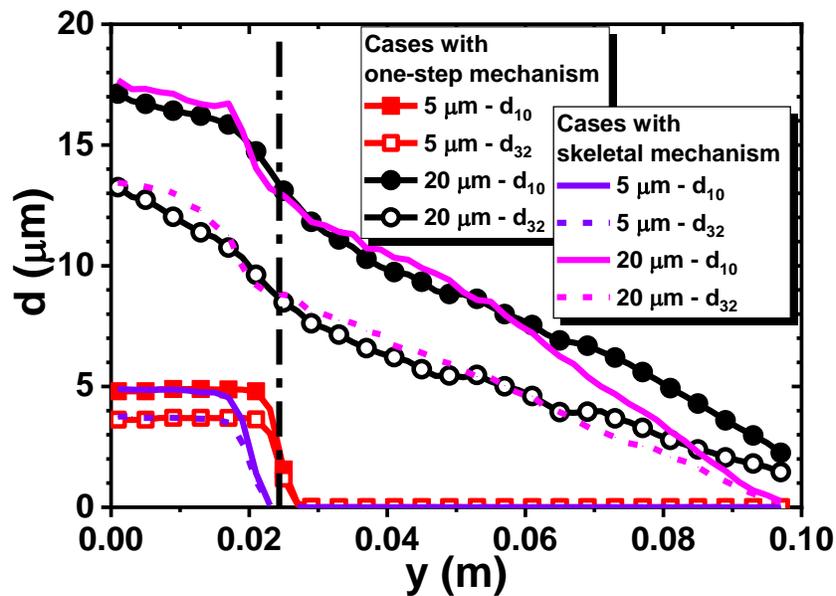

Fig. 17. Profiles of arithmetic mean diameter ($d_{10}$) and Sauter mean diameter ($d_{32}$). Dash-dotted line: mean detonation wave height.

*3.6.3 Droplet aerodynamic breakup*

The effects of the droplet aerodynamic break-up is examined in this section and the ReitzKHRT model [51] is used to simulate the droplet break-up process in high-pressure vaporizing sprays. Figure 18(a) shows the profiles of arithmetic mean diameter $d_{10}$ and Sauter mean diameter $d_{32}$ for Cases 2c ($d_0$ = 20 µm) and 2e ($d_0$ = 80 µm). For Case 2c, it is found that the break-up model would have little effects on the profiles of $d_{10}$ and $d_{32}$. Due to their smallness, limited droplet break-up occurs and the predictions with and without break-up model are not distinguishable. However, for the Case 2e with initial droplets diameter of 80 μm, it can be found that the profiles of $d_{10}$ and $d_{32}$ with the break-up model are slightly less than that without the break-up model at $y < 0.01$ m. In the region 0.01 m $< y <$ 0.024 m (less than the detonation height), the difference of the mean droplet diameters gradually increases. Beyond the mean detonation height (i.e. $y > 0.024$ m), the difference remains almost constant. This is because droplet break-up mainly proceeds immediately behind the detonation front, where large velocity differences between the liquid droplets and the gas phase exist [52]. In spite of the foregoing distinctions, the trends of the mean diameter along the RDE chamber height do not change.

Furthermore, the detonated fuel fractions are also compared over a range of initial droplet diameter in Fig. 18(b). Here the total equivalence ratio is $\phi_t = 1$ and the pre-vaporized gas equivalence ratio is $\phi_g = 0.6$. It is found that the detonated fuel fraction is slightly lower when the droplet break-up model is considered, and the maximum difference (about 2.4%) occurs when the initial droplet diameter is 80 μm. However, in general, the droplet break-up effects on the detonated fuel fraction is marginal. In addition, based on our simulations, the mean propagation speed is negligibly affected by droplet break-up. For instance, the RDW propagation speed for Case 2e without break-up model is 1,501 m/s, which is very close to that predicted with the break-up model, i.e. 1,507 m/s.

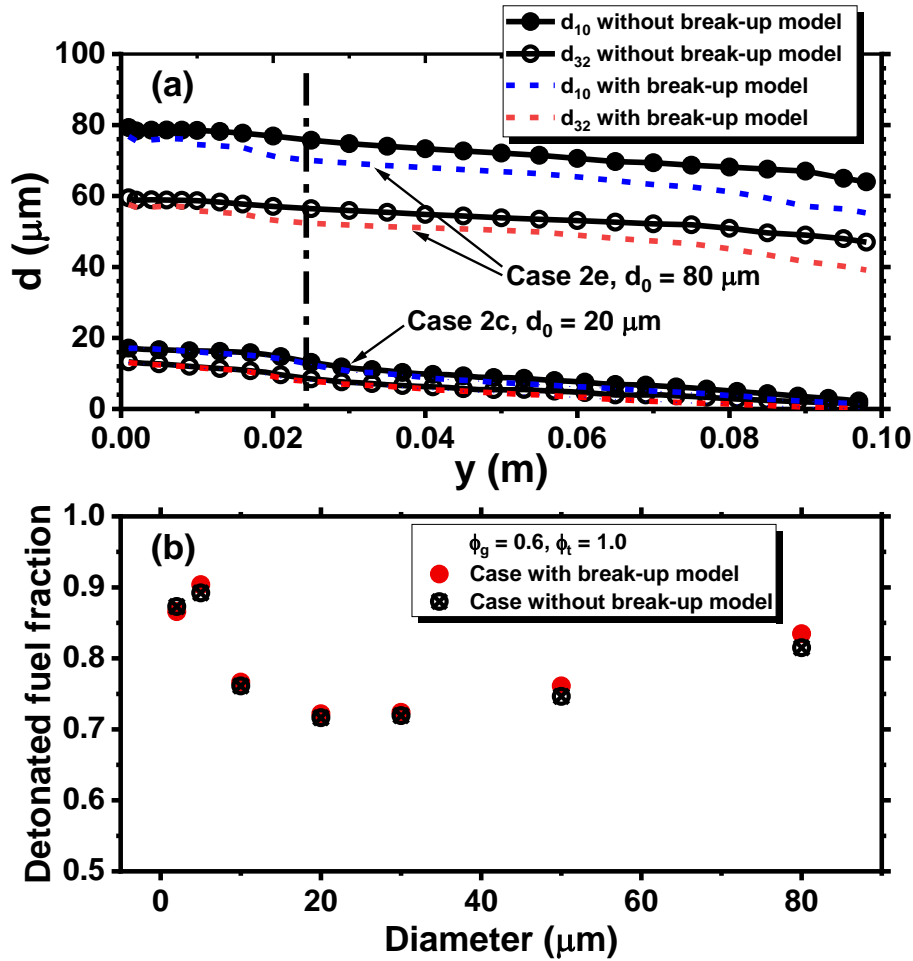

Fig. 18. (a) Profiles of droplet diameter $d_{10}$ and $d_{32}$ along the RDE height with and without break-up model. (b) Detonated fuel fraction as a function of initial droplet diameter with and without droplet break-up model. Dash-dotted line: mean detonation wave height.

## 4. Conclusions

Two-dimensional RDE's with partially pre-vaporized *n*-heptane sprays are simulated with Eulerian−Lagrangian method. Emphasis is laid on the influences of droplet diameter and total (or liquid phase) equivalence ratio on rotating detonation combustion.

The results show that the small *n*-heptane droplets are completely consumed around the detonation wave, while for larger ones, the droplets can cross and continue evaporating behind the detonation wave. Mixed premixed and non-premixed combustion modes exist in the two-phase RDC. Moreover, for large *n*-heptane droplets, the *n*-heptane vapor evaporated from these droplets mixes and reacts with the oxidant behind the detonation wave due to the high temperature, leading to a distribution of spotty diffusion flames.

The detonation propagation speed decreases with increased droplet diameter, and beyond a critical diameter (about 30 µm), it almost does not change with the droplet diameter. The detonated fuel fraction first decreases and then increases with the droplet diameter. It is also found that the detonation propagation speed and detonated fuel fraction change considerably with the pre-vaporized gas equivalence ratio. With increased droplet diameter, the specific impulse first decreases for cases with $d_0$ < 5 µm, then increases with $d_0$ between 5 µm and 20 µm, and finally decreases with $d_0$ > 20 µm. For the cases with small droplets ($d_0$ < 5 µm), decreasing of specific impulse with droplet diameter is mainly because of interphase interaction near the detonation front. However, for the cases with larger droplets ($d_0$ > 5 µm), the change of specific impulse is mainly due to the deflagrative combustion affected by the droplet evaporation behind the detonation wave. Moreover, the droplet distributions in the RDE combustor are considerably affected by the droplet evaporation behaviors.

Finally, sensitivity analysis of mesh resolution, reaction mechanism and droplet break-up model is performed and it is shown that they do not influence the predictions of major RDC features, including rotating detonation propagation, detonated fuel fraction and droplet diameter distribution.

# Acknowledgement

The simulations used the ASPIRE 1 Cluster from National Supercomputing Centre, Singapore (NSCC) (https://www.nscc.sg/). This work is supported by MOE Tier 1 research grant (R-265-000-653-114).

# Data Availability

The data that support the findings of this study are available from the corresponding author upon reasonable request.